\documentclass[letterpaper,prl,twocolumn,superscriptaddress,showpacs]{revtex4}
\usepackage{bm,graphicx}
\usepackage{amsmath}
\usepackage{amsbsy}
\usepackage{amssymb}
\usepackage{array}

\begin{document}

\title{Spin-Wave Relaxation in Diluted Magnetic Semiconductors within the Self-Consistent Green's Function Approach}
\author{J. E. Bunder}
\affiliation{Physics Division, National Center for Theoretical Sciences, Hsinchu 300, Taiwan}
\author{Shih-Jye Sun}
\affiliation{Department of Applied Physics, National University of
Kaoshiung, Kaoshiung 800, Taiwan}
\author{Hsiu-Hau Lin}
\affiliation{Department of Physics, National Tsing-Hua University, Hsinchu 300, Taiwan}
\affiliation{Physics Division, National Center for Theoretical Sciences, Hsinchu 300, Taiwan}
\date{\today}

\begin{abstract}
We employ a self-consistent Green's function approach to investigate the spin-wave relaxation $\Gamma(p)$ in diluted magnetic semiconductors. We find the trend of the spin-wave relaxation strongly depends on the ratio of the itinerant and impurity spin densities. For density ratios in the Ruderman-Kittel-Kasuya-Yosida phase, $\Gamma(p)$ decreases even though thermal fluctuations increase. On the other hand, in the strong coupling phase,  an interesting peak structure appears. We discuss the implications of our numerical results for experiments. 
\end{abstract}

\maketitle

Carrier-mediated ferromagnetism, found in diluted magnetic semiconductors (DMS) such as III-V semiconductors doped with transition metals, opens up the possibility of electric manipulations of magnetic and optical properties because of the exchange coupling between the localized moments of the transition metal and the itinerant spins in the semiconducting bands~\cite{MacDonald05,Wolf01}. In recent years, experimental progress has stimulated intense theoretical investigations of DMS~\cite{Akai98,Dietl00,Konig00,Schliemann01,Litvinov01,Priour04}, which mainly focus on estimates of the critical temperature. Although some experimental data is available~\cite{Crooker96, Akimoto97}, little theoretical attention has been paid to spin dynamics, which is crucially important in understanding the spin coherence. 

\begin{figure}
\rotatebox{270}{\includegraphics[width=4.5cm]{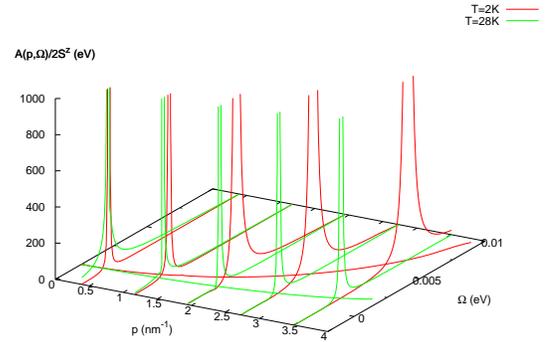}}
\caption{\label{Fig1}
Spin spectral function $A(p,\Omega)/(2\langle S^z\rangle)$ at temperatures $T= 2, 28 K$ with density ratio $n_{h}/n_{I}=0.3$. Against common intuition, the half-width shrinks as the temperature increases.}
\end{figure}

To analyze how spin waves relax in DMS, both the itinerant and impurity spins need to be treated on an equal footing. In this Letter, we employ the self-consistent Green's function
approach\cite{Yang01,Konig01,Bouzerar02,Sun04b,Bouzerar05,Sun06}, which retains both spatial and thermal fluctuations, to investigate the spin-wave relaxation at finite temperatures. The Green's functions allow us to numerically compute the spin spectral function and therefore the spin-wave relaxation. One may naively expect a smooth increase in the spin-wave relaxation as thermal fluctuations grow. However, our numerical results go against common intuition. For instance, in the Ruderman-Kittel-Kasuya-Yosida (RKKY) phase, the spin relaxation, shown in Fig.~\ref{Fig1}, actually decreases with increasing temperatures.

To study the temperature dependence of the spin-wave relaxation, we begin with the simple Zener model,
\begin{equation}
H=H_0+J\int d^3 r\: \bm{S}(r) \cdot \bm{\sigma}(r),
\end{equation}
where $\bm{S}(r)$ and $\bm{\sigma}(r)$ are the impurity and itinerant spin densities, coupled by the antiferromagnetic ($J>0$) exchange coupling. For simplicity, the kinetic energy of the itinerant carriers $H_0$ is assumed to be parabolic, $\epsilon_p = p^2/2m^*$. We take the typical values\cite{Ohno98} for the exchange coupling $J=150$ meV nm$^{3}$ and the impurity spin density $n_{I}=1$ nm$^{-3}$ but vary the ratio of itinerant to impurity spin densities $n_{h}/n_{I}$ in the following.

We now introduce the self-consistent Green's function approach. Note that, instead of treating spin waves as bosons, the Green's function approach respects the spin kinematics which is crucially important when thermal fluctuations are not small. The thermal Green's function, describing the (impurity) spin-wave propagation, is defined as
$
iD(r_1, r_2;t)\equiv  \big\langle \big\langle
T\big[S^+(r_1,t) S^-(r_2,0)\big] \big\rangle \big\rangle,
$
where $T$ is the time ordering operator and the double bracket implies both quantum and thermal averages. To complete the self-consistency, it is necessary to introduce another Green's function which describes the correlations between the impurity and the itinerant spins,
$
iF(r_1, r'_1, r_2;t) \equiv  \big\langle\big\langle T\big[
\psi^{\dag}_{\uparrow}(r_1,t) \psi^{}_{\downarrow}(r_1', t) S^{-}(r_2, 0)
\big] \big\rangle \big\rangle.
$
To simplify the calculations, we assume that the disorder is not strong, i.e. not close to the percolation threshold, and the virtual crystal approximation is valid. Therefore, after a coarse-graining procedure, the translational invariance is approximately restored and the Green's functions only depend on the relative distance. This approximation is justified by recent numerical calculations~\cite{Schliemann01a} which found that magnetization curves with different disorder configurations are almost identical.

\begin{figure}
\rotatebox{270}{\includegraphics[width=4.5cm]{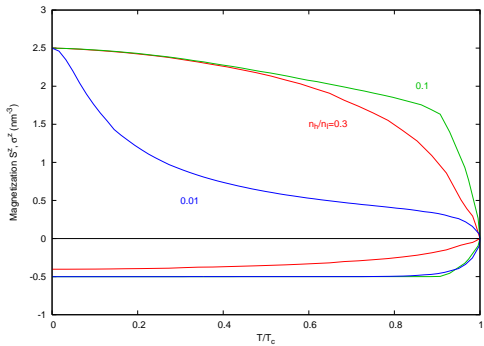}}
\caption{\label{Fig2}
The magnetization curves of impurity and itinerant spins at different density ratios $n_{h}/n_{I} = 0.01, 0.1, 0.3$.}
\end{figure}

The dynamical equations for the Green's functions can be derived by the standard approach and higher-order Green's functions may be decoupled within the random phase approximation. Making use of the translational invariance and the decoupling scheme, the dynamical equation for $D(r;t)$ in the Fourier space is
\begin{eqnarray}
\Omega D(p,\Omega) &=& 2 \langle S^z \rangle - J \langle \sigma^z \rangle D(p,\Omega)
\nonumber\\
&+& J \langle S^z \rangle \int \frac{d^3k}{(2\pi)^3} F(k,k+p;\Omega),
\label{D}
\end{eqnarray}
where $F(k,k+p;\Omega)$ is the Fourier transformed Green's function $F(r,r';t)$. Similarly, the other dynamical equation is $F(k, k+p;\Omega)= G (k, k+p;\Omega) D(p,\Omega)$ with
\begin{equation}
G (k, k+p;\Omega)=\frac{J}{2}\frac{ f_{\uparrow}(\epsilon_k)- f_{\downarrow}(\epsilon_{k+p})}{\Omega+\epsilon_k-
\epsilon_{k+p}+\Delta+i\gamma_{h}},
\label{eq:F-D}
\end{equation}
where $f_{\alpha}(\epsilon_k)=
[e^{\beta(\epsilon_k+\alpha\Delta/2-\mu)}-1]^{-1}$ is the Fermi distribution function for itinerant carriers with the Zeeman gap $\Delta=J\langle S_z\rangle$. On substituting the expression for $F(k,k+p;\Omega)$ into Eq.~(\ref{D}), the spin-wave propagator is solved,
\begin{eqnarray}
D(p,\Omega)&= & \frac{2\langle S_z\rangle}{\Omega-\Sigma(p,\Omega)+i \gamma_{I}},\label{eq:D2}
\end{eqnarray}
where the prefactor $2\langle S^z \rangle$ comes from the exact treatment of spin-wave kinematics.  The self-energy $\Sigma(p,\Omega)$ arises from interactions between the itinerant and the impurity spins and contains two terms,
\begin{eqnarray}
\Sigma(p,\Omega) = -J \langle\sigma^z\rangle
+\Delta
\int \frac{d^3k}{(2\pi)^3}G(k,k+p;\Omega).
\label{eq:gf}
\end{eqnarray} 
If only the first term is retained, it coincides with the Weiss mean-field approximation. However, since the first term is always real, dropping the second term would remove all information about the Landau damping of the spin waves, described by the imaginary part of the self-energy $\Sigma_{I}(p,\Omega)$, inside the Stoner continuum.

Once the spin-wave dispersion is obtained from $\omega_p -\Sigma_{R}(p,\omega_p)=0$, the polarization of the impurity spins can be computed by Callen's formula~\cite{Callen63},
\begin{equation}
\frac{\langle S^{z}\rangle}{n_{I}}=S-\langle n_{sw}\rangle +\frac{(2S+1)\langle n_{sw}\rangle^{2S+1}}{(1+\langle n_{sw}\rangle)^{2S+1}-\langle n_{sw}\rangle^{2S+1}}.
\label{eq:sz}
\end{equation}
where the average number of spin waves is $ \langle n_{sw}\rangle=(1/n_I)\int d^3 k/(2\pi)^3 \:\:[e^{\beta\omega_k}-1]^{-1}$. The difference between the independent spin-wave theory and the self-consistent Green's function method lies in the third term, which correctly accounts for the spin kinematics. The magnetization curves for different density ratios $n_{h}/n_{I}$ are shown in Fig.~\ref{Fig2}. While this is not the main focus of our results, we emphasize the sensitivity in the shape of the magnetization curve to the density ratio. From previous studies in the diffusive regime, there are three phases for the Zener model -- mean-field, RKKY and strong coupling. For $n_h/n_I=0.3$, it belongs to the RKKY phase and the magnetization curve is rather conventional. For $n_h/n_I=0.1$, it lies in the strong coupling phase and the magnetization curve decreases linearly with a very small gradient and, only in the narrow regime near the critical temperature, the magnetization dives to zero and the phase transition occurs. Similar magnetization curves, quite different to what we expect for a Heisenberg-like model, have been observed in some experiments~\cite{Edmonds04}. Finally, we choose $n_h/n_I=0.01$ (also in the strong coupling phase) to demonstrate that the magnetization curve can turn {\em concave}, though this is not necessarily related to percolation near the metal-insulator transition.

\begin{figure}
\rotatebox{270}{\includegraphics[width=4.5cm]{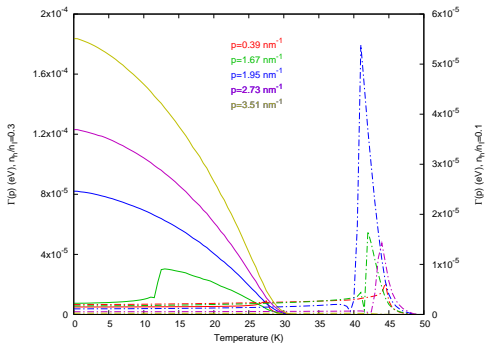}}
\caption{\label{Fig3}
Spin-wave relaxation rate $\Gamma(p)$ at different momenta and density ratios $n_{h}/n_{I} = 0.3, 0.1$. The solid (broken) lines correspond to the left (right) axis.}
\end{figure}

The unique feature of the Green's function approach is that we can also study the temperature evolution of the spin spectral function, $A(p,\omega) = -(1/\pi) \mbox{Im} D(p,\Omega)$, as shown in Fig.~\ref{Fig1}. From Eq.~(\ref{eq:D2}), the imaginary part of the spin-wave propagator takes the Lorentzian form,
\begin{eqnarray}
A(p,\omega) = 2\langle S^z \rangle \left(\frac{Z}{\pi}\right) \frac{\Gamma(p)}{(\Omega-\omega_p)^2 + \Gamma(p)^2},
\label{eq:imagD}
\end{eqnarray}
where $Z \approx 1$ is the spectral weight for the gapless spin wave. After Taylor expansion of $\Sigma_I(p,\Omega)$ in the vicinity of the spin-wave dispersion $\Omega = \omega_p$, the relaxation rate of the spin waves $\Gamma(p) \approx \Sigma_I(\omega_p)$. The numerical results for the spin-wave relaxation rate versus temperature with $n_{h}/n_{I}=0.3, 0.1$ are shown in Fig.~\ref{Fig3}. One may expect that increasing thermal fluctuations lead to a increasing relaxation rate, but our numerical results disagree. In the RKKY phase with $n_{h}/n_{I}=0.3$, the spin relaxation rate $\Gamma(p)$ decreases as the temperature increases. This reduction can be understood as the weakening of the effective coupling strength between the itinerant and impurity spins in proportion to the spin polarization. However, since the decoupling scheme in the self-consistent approach only includes decaying channels in the diagonal parts (as in all random phase approximations), the computation of $\Gamma(p)$ breaks down in the vicinity of the critical temperature where off-diagonal channels dominate. Therefore, one should not take the vanishing $\Gamma(p)$ at the Curie temperature seriously. However, in the low temperature regime, the diagonal channels do dominate and our numerical result of decreasing $\Gamma(p)$ with increasing temperature is a real effect. 

In the strong coupling phase with $n_{h}/n_{I}=0.1$, the trend of the spin relaxation is completely different. The spin relaxation $\Gamma(p)$ is small and does not change much at low temperatures, then a peak structure appears just before hitting the critical temperature. This non-monotonic trend can be explained by the energy shift of the Stoner continuum. In the strong coupling phase, the Stoner continuum is separated from the gapless spin waves at low temperatures by a finite gap. The gap suppresses $\Gamma(p)$ and makes it insensitive to temperature changes. The gap disappears at the temperature where the polarization of the itinerant spins $\langle \sigma^z \rangle$ is no longer fully polarized. Due to the intersection of the spin-wave dispersion and the Stoner continuum, enhanced spin relaxation is expected. An analysis of our numerical results reveal that the partial polarization of $\langle \sigma^z \rangle$ and the peak in $\Gamma(p)$ occur at the same temperature. 

One may notice the close relationship between the magnetization curve and the spin relaxation rate and their sensitive dependence on the density ratio $n_h/n_I$. This is reminiscent of the interesting differences between as-grown samples and annealed ones\cite{MacDonald05}. The annealing process removes interstitial impurities and reduces the disorder strength. However, if the as-grown sample is already in the diffusive regime, the annealed one can also be described by the virtual crystal approximation. We believe the major difference  arises from the increase in carrier density and thus the density ratio $n_h/n_I$. The annealing process increases $n_h/n_I$ from the strong coupling to the RKKY phase which changes the shape of the magnetization curve. Therefore, it would be exciting to study the spin dynamics for both as-grown and annealed samples and compare the experimental outcomes with our numerical predictions. Note that we did not include the realistic six-band Luttinger model here. In principle, our approach can be generalized to include more bands by introducing more Green's functions but the self-consistent equations will be rather complicated, not fatal though. However, it is important to emphasize that the ratio $n_h/n_I$ to enter the RKKY regime is expected to be larger, as compared to the estimate from the two-band model.

We thank Sankar Das Sarma and Allan MacDonald for fruitful discussions. HHL acknowledges supports from National Science Council in Taiwan through grants NSC-93-2120-M-007-005 and NSC-94-2112-M-007-031 and also supports from KITP by the National Science Foundation under Grant No. PHY99-07949.

\end{document}